\documentclass[12pt]{article}
\usepackage[dvips]{graphicx}
 \newcommand{\cc}{\cite}

\newcommand{\be}{\begin{equation}}
\newcommand{\ee}{\end{equation}}

\def\ve{\varepsilon}
\def\w{\omega}

\def\pd{\partial}

\def\L{\Lambda}

\def\ex{\hbox{e}}

\def\<{\langle}
\def\>{\rangle}

\def\Log{\hbox{ln}}
\def\a{\alpha}
\def\b{\beta}
\def\g{\gamma}  \def\G{\Gamma}
\def\d{\delta}  \def\D{\Delta}
\def\l{\lambda}   \def\L{\Lambda}
\def\s{\sigma}
\def\r{\rho}  
\def\x{\xi}
\def\c{\chi}

\def\m{\mu}
\def\n{\nu}
\def\t{\tau}

\def\w{\omega}

\def\v{\vec}

\def\vf{\varphi}
\def\({\left(}
\def\[{\left[}
\def\){\right)}
\def\]{\right]}

\def\cot{\hbox{cot}}
\def\cos{\hbox{cos}}
\def\sin{\hbox{sin}}

\def\pd{\partial}
\def\dk{{d^n k \over (2\pi)^n}}
\def\Tr{\hbox{Tr}}
\def\pa{{\cal P}}
\def\w1{W^{(1)}}
\def\v1{V^{(1)}}
\def\prop{D_{\m\n}}

\def\dI{\int \! dn(\r)}

\def\tim{\lambda}
\begin{document}
\begin{center}
{\large \bf{Instanton Contribution to the Quark Form Factor}} \\
\vspace{10mm}
{\small Alexander E. Dorokhov$^{a,}$\footnote{dorokhov@thsun1.jinr.ru} {\footnotesize and} Igor O.
Cherednikov$^{a,b,}$\footnote{Igor.Cherednikov@jinr.ru}}
\\
 \vspace{4mm}
 $^a$ {\small \it Bogolyubov Laboratory of Theoretical Physics,
Joint Institute for Nuclear Research \\
141980 Dubna, Russia}\\
\vspace{3mm}
$^b$ {\small \it Institute for Theoretical Problems of Microphysics,
Moscow State University \\ 119899 Moscow, Russia}
\end{center}
\begin{center}
{\small (17 Oct 2002)} \\
\end{center}

\vspace{5mm}

\begin{abstract}
\noindent
{\small The nonperturbative effects in the quark form factor
are considered in the Wilson loop formalism.
The properties of the Wilson loops with cusp
singularities are studied  taking into account the perturbative and
nonperturbative contributions, where the latter are considered
within the framework of the instanton liquid model.
For the integration path corresponding to this form factor --
the angle with infinite sides -- the explicit expression for
the vacuum expectation value of the Wilson operator is found to leading order.
The calculations are performed in the weak-field limit for the instanton vacuum contribution
and compared with the one- and two-loop order results for the perturbative
part. It is shown that the instantons produce the powerlike corrections to
the perturbative result, which are comparable in magnitude with the perturbative part at the
scale of order of the inverse average instanton size.
It is demonstrated that the instanton contributions to the quark form factor
are exponentiated to high orders in the small instanton density parameter.}
\end{abstract}

\vspace{5mm}
\section{Introduction}
\noindent
A convenient approach which may be applied to
the investigation of the infrared behavior of QCD is the formulation of the non-Abelian
gauge theory on loop space that becomes possible by virtue of
the close correspondence between gauge and chiral fields  \cc{pol}.
One of the advantages of this method is that it allows one to consider
both the perturbative and nonperturbative contributions.
The basic object of study in such an approach is the gauge invariant
vacuum average of the Wilson loop operator \be
W(C) = {1 \over N_c} \Tr \<0|  \pa \exp \( i g \int_C \! d x_{\m} \hat A_{\m}
(x) \)|0\> \ , \label{1} \ee where
the integration goes along the closed contour $C$ and the gauge field
\be
\hat A_{\m} (x) = T^a A^a_{\m}(x)\ , \   \ T^a = {\lambda^a
\over 2} \ \ ,  \ee belongs to the Lie algebra of the gauge group
$SU(N_c)$, while the Wilson loop operator $\pa \ex^{ig\int\! dx A(x)}$ lies (for quark lines) in
its fundamental representation.

In particular, the objects such as  path-ordered contour integrals of the gauge
field emerge naturally in the study of hard collisions of hadrons (Drell--Yan
process, deep inelastic scattering, etc.) and the hadron form factors in
the eikonal approximation \cc{hard, nach}. The use of the Wilson integral
formalism in the heavy quarks effective theory has also been discussed (see, {\it e.g.,}
\cc{korh}). In the case of hard processes, the integration contours
become infinite, since they present the classical trajectories of the partons
participating in the collision.
Considering the parton scattering processes at large total energies and small
momentum transfers it is  revealed that
the role of the resumed soft gluons exchanges becomes significant.
While in QED this problem is solved in closed form yielding the well-known
Sudakov form factor (by virtue of exponentiation of the leading one loop
perturbative term), in QCD the situation appears to be more complicated due to its
non-Abelian nature and the influence of original nonperturbative effects.
Nevertheless, it has been shown in perturbation theory that in QCD processes the exponentiation
takes place with the exponent given by a series in the strong coupling
constant. Additional soft contributions may appear due to various nonperturbative
effects, a number of which are closely related to the nontrivial structure of
the QCD vacuum. In the case of the instanton ensemble modeling the QCD vacuum,
the exponentiation of the single instanton contribution can also be proven
(see Section 4).
Indeed, although the QCD vacuum plays an important role in the high-energy
collisions, the direct investigation of these effects remains a difficult task.
The aim of the present work is to evaluate them within the framework of the instanton liquid
model of QCD vacuum for the simplest possible quantity -- the quark form factor.

It is generally believed that the nonperturbative structure of QCD can be well
understood within the instanton liquid model \cc{rev}. Considering the
QCD vacuum as an ensemble of instantons, one can describe a lot of
the low-energy phenomena in strong interactions on the qualitative, as well as
the quantitative level. The importance of the instanton induced effects of strong
interactions is also supported by lattice simulations \cc{rev, lat}. The instanton
picture is generally considered as a fruitful and perspective framework for
hadronic physics. The role of the instantons in hard hadronic processes has
been studied intensively, both theoretically and experimentally. The perspectives for an
unambiguous experimental detection of instanton contributions are believed to be
optimistic and very promising. The main problem arising in this context is the evaluation
of instanton induced effects at a hard scale of a given process \cc{ring}.
Hence, we are going to study the instanton
induced contributions to the Wilson integral over contours of a various
geometry and apply the obtained results to the analysis of nonperturbative
effects in the high-energy hadron collisions. Recently a similar approach has been
developed in the study of high energy elastic and quasielastic parton-parton scattering \cc{sh1}.
The vacuum averages of Wilson operators with finite-length contours were considered
in the instanton model in \cc{DEMM99}, where the properties of the nonlocal quark and
gluon condensates were discussed.

In the present paper, we start the investigation of the
instanton induced effects in the high-energy QCD processes. We propose an
approach which allows one to evaluate the instanton contributions to the Wilson
integrals made of several (in)finite lines containing specific cusp and/or cross
singularities. For this purpose, we start with one of the simplest
configurations, {\it i. e.,} the angle with infinite sides that corresponds
to the integration path for the Wilson operator describing the soft part of
the quark Sudakov form factor \cc{kr217}.
Such configurations enter as constructive units in the Wilson
contours for many high energy processes.

This paper is organized as follows.
In Section 2, we briefly describe the general renormalization properties
of the Wilson loops with cusp singularities and find the explicit formula
in leading order, applicable both to perturbative and nonperturbative fields.
Then we analyze the UV and IR divergencies  and calculate the perturbative part
in an arbitrary covariant gauge demonstrating explicitly the gauge invariance
of the obtained (renormalized) result. Section 3 is devoted to the
study of the instanton induced nonperturbative contributions. The
expression for them is found in the closed form in the weak-field limit in terms of the instanton
profile function. It is shown that the instantons actually give powerlike corrections to the
perturbative result. The magnitude of the instanton induced effects is found
to be comparable with the perturbative part at a certain low momentum scale.
In Section 4, we prove the exponentiation in the small instanton density parameter
of the all-order single instanton contribution to the Sudakov form factor.
Afterwards, we discuss the possible applications of the results obtained above to investigations of
the physically interesting processes and make some conclusive remarks.
\vspace{4mm}

\section{Renormalization and perturbative contribution}
\noindent
Let us consider in short the renormalization properties of the Wilson loop.
It is well known, that evaluation of $W(C)$ (\ref{1}) encounters divergencies.
The divergencies appearing for smooth contours without
self-intersections can be removed by means of the convenient  R operation
\cc{ao, bra}:  \be W_R(C, g_R, \m) = \lim_{\ve \to 0 } \widetilde W (C, g_R,
\m, \ve) \equiv \lim_{\ve \to 0} R W (C, g, \ve) \ ,   \ee where $\ve$ is
the dimensional regularization parameter, $\mu$ is the UV
normalization scale.  In this case, the R operation consists in the multiplicative
renormalization of the coupling constant $g \to g_R$  and the field
$\hat A_\m \to \hat A_\m^R$.

In the more general case, the contour may contain a
number of cusps, {\it i. e.,} the points where the derivative is not smooth.
In this situation,  divergencies of another type arise which depend on the
corresponding cusp angles $\g_i$. Although they were considered
as a shortcoming of the loop formulation of a gauge theory about two decades
ago, these singularities (and the anomalous dimensions related to them),
are shown to play an important role in the partonic hard processes
controlling the asymptotics of the scattering amplitudes \cc{kr325}.
In order to remove these divergencies, the generalized $K_\g$ operation has been
proposed \cc{bra}. For a contour with one cusp $C_\gamma$ this operation is reduced to
  \be W_R(C_\g, g_R, \m, \overline{C}_\g) = \lim_{\ve \to 0 } K_\g
\widetilde W (C_\g, g_R, \m, \ve)\ ,
\ee where
the contour $\overline{C}_\g$ is treated as a subtraction point.
It has been proved that the $K_\g R$ operation renormalizes
multiplicatively any loop integral with finite number of cusps \cc{bra, kr1}.
The multiplicative renormalizability means that the $K_\g$ operation acts as
\be
K_\g \widetilde W (C_\g, g_R, \m, \ve) = Z_{cusp}(g_R, \g, \m, \ve)
\widetilde W (C_\g, g_R, \m, \ve) \ ,
\ee where
\be
Z_{cusp} = \widetilde W^{-1}(\overline{C}_\g, g_R, \m, \ve)
= 1 + \sum_{n=1}^{\infty} \ \({\a_S \over \pi}\)^n Z_n (\g, \m, \ve) \
\ee is an additional renormalization constant. The coefficients
$Z_n$ are the sum of simple poles plus some finite terms
\be
Z_n(\g, \ve) = \sum_{k=1}^n \ \( {1 \over \ve^k} A_{kn} (\g)\) + (\hbox{finite \ \  terms})_n \ .
\ee
A choice of the finite part of $Z_n$ to each order actually determines the
$K_\g$ subtraction scheme.
In order to remove pole terms we shall  use the modified minimum subtraction
${\overline{MS}}$-scheme $(K_\g^{\overline{MS}})$.

The first nontrivial cusp dependent term in the expansion of $W(C)$ (1) in
powers of $g^2$ for the angle with two infinite straight line rays
[(Fig. 1a)] contains the contributions from both
perturbative [(Fig. 1b)] and nonperturbative [(Fig. 1c)] fields
\be
\w1 (\g) = \w1_P(\g) + \w1_{NP}(\g)  \ . \label{pnp}
\ee

Regardless of which part is considered -- either perturbative, or
nonperturbative -- we will write the one-loop contribution in the following form:
\be
\w1 (\g) =  -{g^2 C_F \over 2}  \ \int_{C_\g}\! dx_\m \int_{C_\g}\! dy_\n \ \prop (x-y)
\ , \label{g1} \ee where $C_F= {N_c^2 -1 \over 2N_c}$.
It is convenient to present the gluon propagator $\prop(z)$ in the form
\be
\prop(z) = \d_{\m\n}
\pd_z^2 d_1(z^2) - \pd_\m\pd_\n d_2(z^2) \ .
\label{st1} \ee
Here and in what follows, we use the dimensional regularization with $n= 4 -
2\ve$, $\ve <0 $ in order to control the IR-divergent terms in the
integrals. The remaining UV singularity (due to the infinitely small $z^2$ in the vicinity of the
cusp) will be regularized by the corresponding UV cutoff.

The trajectories of the incoming and outgoing quarks [(Fig. 1a)] may be
parametrized as  $$ x = v_1 s \  (0 < s < \infty) \ \ , \ \ y = v_2 \t \
(-\infty < \t < 0) \ . $$
The angle between the vectors $v_1$ and $v_2$ is
given in the Minkowski space by
\be \cosh \c = (v_1 v_2) = {(p_1 p_2) \over m^2} = 1 + {Q^2 \over 2 m^2} \ \ ,
\ \ - Q^2 = (p_2 - p_1)^2 \ \ , \ \ v_{1,2}^2 = 1,   \label{m1} \ee
where the quark momenta are supposed to be on shell: $p_1^2=p_2^2=m^2$.
The continuation to the
Euclidean space is defined as \cc{kr1, meg}:
\be
\c \to i\g\ . \ee By virtue of the interpretation of the graph Fig. 1 as an
amplitude of the elastic scattering of an on-mass-shell one-dimensional
fermion on a color singlet potential \cc{ao, kr1}, we have to consider the
quantity
\be \w1 (\g) = \widetilde \w1(\g) - \widetilde \w1(0) \ ,
\label{se1} \ee where (for some technical details of the calculations, see
the Appendix)
\be
\widetilde \w1(\g) = - g^2 C_F \[ (n-2)d_1(0) \g \cot \g  + d_2(0) \] \ .  \ee
It follows from Eq. (\ref{se1}) that the integrals (\ref{g1}) in which
both points $x$ and $y$  belong to the same side of the angle do not contribute
to the quantity $\w1 (\g)$.
Hence we have within the one loop accuracy
\be W (\g) = 1  - 4\pi \a_S C_F (n-2) h(\g) d_1(0) \ ,
 \label{pe1} \ee
where
\be
h(\g) = \g \cot \g-1
\ee
is the universal cusp factor.
We should emphasize here that the expression (\ref{pe1}) holds for
perturbative part as well as for the nonperturbative part depending on the value $d_1(0)$. For the
perturbative field, Eq. (\ref{pe1}) reflects the explicit gauge invariance in the set of
covariant gauges, since the gauge fixing parameter $\xi$ enters only in the function $d_2(z^2)$.

Let us consider first the perturbative part $\w1_P(\g)$.
By using the free propagator in the Euclidean space (see the Appendix) the IR-regularized
value of $d_1(0)$ can be written in the form
\be d_1(0; \ve, \l) =  {(\l^2\pi)^\ve \over 16
\pi^2} \int_0^{\infty} \! d\a \a^{-(1+\ve)} \ , \ee
where $\l^2$ is the IR regularization parameter.
This integral diverges at the
upper (UV) limit for $\ve < 0 $, and hence we must regularize it.
To this end, we may introduce the UV cutoff $\m^2$, that corresponds to the following
replacement in the
denominator of the perturbative propagator: $z^2 \to z^2+\m^{-2}$, and finally we get
\be d_1(0; \ve, \m/\l) = - {1 \over \ve } {1 \over 16 \pi^2}\({\l^2\pi \over \m^2}\)^\ve \ .
\ee
Thus one obtains, with one loop accuracy,
\be
W_P (\g, \a_S, \m/\l, \ve) = 1 + {\a_S \over 2 \pi} C_F h(\g)
\ {1 \over \ve } (1-\ve) \({\l^2 \pi \over \m^2}\)^\ve \ , \label{z1}
\ee and the cusp renormalization constant $Z_{cusp}^{(1)}$ within the ${\overline{MS}}$-scheme:
\be
Z_{cusp}^{(1)} (\g, \ve) = 1 - {\a_S \over 2 \pi} C_F h(\g) \ \({1 \over \ve } + \Log \pi -1\)\ ,
\ee which is in agreement with \cc{hard0, bra}. Therefore, the perturbative part of the finite
renormalized function reads
\be W_P(\g, \m) = \lim_{\ve \to 0}Z^{(1)}_{cusp} (\g, \m, \ve)
\widetilde W_P (\g, \m, \ve) =1 - {\a_S \over 2 \pi} C_F h(\g) \Log {\m^2 \over \l^2}
\ . \label{5}\ee
The one-loop cusp anomalous dimension which satisfies the RG equation reads
\be
\(\m { \pd \over \pd  \m} + \b(g_R){\pd \over \pd g_R} \) W_P (\g, \m, \a_S(\m)) = -
{\G^{cusp}}^{(1)}_P (\g, \a_S(\m)) W_P (\g, \m, \a_S(\m))  \ , \label{01} \ee
 where $\b(g_R)$ is the usual QCD $\b$-function
$\b(g_R) = \m {\pd \over \pd\m} g_R = \hbox{const} \   g_R^3 + O(g_R^5)$,
which can be found
in the one-loop order \cc{hard0, bra}
\be {\G^{cusp}}^{(1)}_P (\g, \a_S) = - {d \over d\Log \m} W^{(1)}_P(\g, \m) =
{\a_S \over \pi} C_F h(\g) \ . \label{cp} \ee
Hence, we reproduce the Wilson operator value for the infinite contour
with the Euclidean cusp parameter $\g$.

\vspace{1cm}

\section{Instanton contribution}
\noindent
Let us estimate the nonperturbative contribution to $W(C)$ in the
instanton model. The instanton field is given by
\be \hat A_\m (x; \r) = A^a_{\m} (x; \r) {\sigma^a \over 2} = {1 \over g}
 {\hbox{\bf R}}^{ab} \sigma^a {\eta^{\pm}}^b_{\m\n} (x-z_0)_\n \vf
(x-z_0; \r) , \label{if1}\ee
where ${\hbox{\bf R}}^{ab}$ is the color orientation matrix $(a,b=1,2,3)$,
$\sigma^a$'s are the Pauli matrices,
and $(\pm)$ corresponds to the instanton or antiinstanton.
The averaging of the Wilson operator over the nonperturbative vacuum is reduced to the integration
over the coordinate of the instanton center $z_0$, the color orientation and the
instanton size $\r$.  The measure for the averaging over the instanton ensemble
reads $dI = d{\hbox{\bf R}} \ d^4 z_0 \ dn(\r) $, where
$ d{\hbox{\bf R}}$ refers to the averaging over color orientation,
and $dn(\r)$ depends on the choice of the instanton size distribution.
Taking into account (\ref{if1}),
we write the Wilson integral (\ref{1}) in the single instanton approximation
in the form
\be
w_I(C) = {1\over N_c}  \<0| \Tr  \exp \( i \sigma^a \phi^a \)|0\> \ ,
\label{wI1}\ee
where the phase is
\be \phi^a =
{\hbox{\bf R}}^{ab} {\eta^{\pm}}^b_{\m\n} \int_{C_\g} \! dx_\m \ (x-z_0)_\n
\vf (x-z_0; \r) \ . \ee We omit the path
ordering operator $\pa$ in (\ref{wI1}) because the instanton field
(\ref{if1}) is a hedgehog in color space, and so it locks the color
orientation by space coordinates. Performing the averaging over the
color rotations, making use of the known property of 't Hooft symbols,
\be
{\eta^{\pm}}^a_{\m\n} {\eta^{\pm}}^a_{\r\s} = \d_{\m\r}\d_{\n\s} -
\d_{\m\s}\d_{\n\r} \pm \ve_{\m\n\r\s} \ ,
\ee
and performing the subtraction of self-energy part (\ref{se1})
we obtain the all-order single instanton contribution to the cusp-dependent part of Wilson loop
(\ref{1}):
\be
w_I(\g) =   \int \! d^4z_0 \int \! dn(\r) \  \[\cos \ \phi (\g, z_0, \r) - \cos \ \phi (0, z_0, \r)  \]\
, \label{it1}
\ee
where the squared phase $\phi^2 = \phi^a\phi^a$ may be written as
$$
\phi^2 (\g, z_0, \r)=  \sum_{i,j=1,2} \[ (v_iv_j) z_0^2   -
(v_iz_0)(v_jz_0)\]
\ \times $$ \be
\int_0^\infty \! d\s \vf\[((-1)^{i+1} v_i\s - z_0)^2; \r \]
\int_0^\infty \! d\s' \vf \[((-1)^{j+1} v_j\s' - z_0)^2; \r \]  \ . \label{it2}
\ee

Although the expressions (\ref{it1}, \ref{it2}) give
the complete formula for the all-order single instanton contribution,
in what follows we restrict ourselves to the investigation of the weak field limit.
Phenomenologically, it is assumed  that the instanton size
distribution is sharply peaked at a certain finite value.
As we will see soon, this finite instanton size $\r$ provides an UV cutoff
for the nonperturbative gluon propagator.
Therefore, for the instanton field we use the same dimensional IR regularization with $\ve <
0$ and the IR parameter $\l$ as in the perturbative case\footnote{
Another possible approach to divergencies in the integrals over the instanton fields is
to use a constrained instanton solution which is characterized by the
profile function exponentially decreasing at large distances  \cc{sh1, DEMM99}.}.
So, in this limit
the leading instanton induced term reads  \be
w^{(1)}_I(\g) = - {g^2 \l^{n-4} \over 2} \ \dI \  \int_{C_\g}  \! dx_\m
\int_{C_\g}  \! dy_\n  \int \! \dk \ \tilde A^a_\m (k; \r)
\tilde A^a_\n (-k; \r) \ex^{-ik(x-y)} \ . \label{i1} \ee
By using the Fourier transform of the instanton field
\be \tilde A^a_\m (k; \r) =  - {2i \over g }  {\eta^\pm}^a_{\m\s}
k_\s \tilde \vf'(k^2; \r ) \ ,  \ee  Eq. (\ref{i1}) can be written in the
form of Eq. (\ref{st1}) with the instantonic analogue of the function $d_1(z^2)$:
\be
d_1(z^2) \to d_1^I(z^2) = - {1 \over g^2 C_F} \int \! dn(\r) D_I(z^2; \r,
\tim)\ , \ee where
\be
D_I (z^2; \r, \l) =  \l^{4-n} \int \! \dk \ex^{-ikz} \[2 \tilde \vf'(k^2; \r)\]^2 \ . \label{i3}
\ee
Above, $\tilde \vf (k^2; \r)$
is the Fourier transform of the instanton profile
function $\vf (z^2; \r)$ and $\tilde \vf'(k^2; \r)$ is it's derivative with respect to
$k^2$.
Now using the result (\ref{pe1}) of the previous section, we get the
instanton contribution in the form
\be w_I(\g; \ve, \tim) = (n-2) h(\g) \dI \ D_I(0; \ve, \tim, \r) \ .\label{i4}\ee

Consider now the renormalization of the nonperturbative part for the instanton field in
the singular gauge,\footnote{
We consider the singular gauge, since in the regular gauges
the profile function decreases insufficiently
rapidly at infinity, and therefore the contribution of the infinitely
distant part of the contour must be taken into account, in contrast to the
singular gauge where this contribution vanishes.}
where the profile function is
\be
\vf(u; \r) = {\r^2 \over z^2(z^2+\r^2)} \ .
\label{SingG}\ee
By using the Fourier transform of this function
one gets the IR dimensionally  regularized $D_I(0; \ve, \tim, \r)$ in the form
$$
D_I(0; \ve, \tim, \r) = $$ \be = {\r^4 \pi^{2+ \ve} \l^{2\ve} \over 4} \  \int_0^{\infty} \! d\a_1 d\a_2
\int_0^{\infty} \! d\b_1
d\b_2 \ {\ex^{-\r^2(\a_1+\b_1)} \over (\a_1+\a_2)^{1+\ve}(\b_1+\b_2)^{1+\ve}
\(\a_1+\a_2+\b_1+\b_2\)^{2-\ve}} \  .
\ee
It is clear from this expression that the instanton size plays a role of the UV cutoff, since it provides
an exponential suppression of the integrand  at large $(\a, \b)$ what corresponds to the large momentum.
Computing this integral we have $$
D_I(0; \ve, \tim, \r) = $$ \be - {\r^4 \pi^2 \over 4} \(\r^2 \tim^2 \pi\)^{\ve} \ {\G(1- \ve) \over  \ve} \
\(1 + \ve \int_0^1 \! \int_0^1 \! \int_0^1 \! dx dy dz \ \Log\[{xz +
y(1-z) \over z(1-z)} \] + O(\ve^2)\) \ .
\ee
Applying the renormalization procedure as described in the previous section, we find
in the leading order the instanton contribution to the renormalization constant
\be
Z_I(\g, \ve) =1 - 2 h(\g) \ \({1 \over \ve } \ \dI \ {\r^4 \pi^{2} \over
4} + \D_{NP} \)\ ,
\ee where the finite term is
\be \D_{NP} =
\g_E +  \Log \pi + {\pi^2 \over 6} -{3 \over 2} \ ,  \ee
and obtain the corresponding contribution to the Wilson loop
\be
w^{(1)}_I(\g, \tim) = 1 +  \pi^2 h(\g) \dI \ \r^4 \Log {\left(\r\tim\right)}\ .  \label{II1}\ee

In order to estimate the magnitude of the instanton induced effect
we consider the  distribution function which has been suggested in
\cc{shdist} (and discussed in \cc{DEMM99} in the framework of constrained instanton
model) in order to describe the lattice data \cc{lat}, namely the following one
\be
dn(\r) = {d\r \over \r^5} \ C_{N_c} \(2\pi \over \a_S(\r) \)^{2N_c} \exp\(- {2\pi \over \a_S (\r)}
\) \exp\(- 2 \pi \s \r^2\) \ , \label{dist1}
\ee where the numerical constant $C_{N_c}$ is determined by the number of
colors
\be
C_{N_c} = {0.466 \ \ex^{-1.679 N_c} \over (N_c-1)! (N_c-2)!}\approx 0.0015 \ ,
\ee
and the string tension is accepted to be $\s \approx (0.44 \
GeV)^2$ \cc{sh1, shdist}.  Then, using the one loop expression for the running
coupling constant
\be
\a_S (\r) = - {2\pi \over \b_0 \Log \r \L_{QCD}} \ \ , \ \ \b_0 = {11 N_c - 2 n_f \over
3}\ \ ,
\ee we find the instanton contribution (\ref{II1}) in the form [in the distribution
(\ref{dist1}), the slow varying logarithmic factor due to the power of the coupling $\a_S$ is
assumed to be constant, and taken at the point of the mean instanton size $\bar
\r$]
\be
w^{(1)}_I(\g, \tim) =
1 + \pi^2  h(\g) {C_{N_c} \G (\b_0/2) \over 4} \(2\pi \over \a_S(\bar \r) \)^{2N_c}
\({\L_{QCD} \over \sqrt{2\pi
\s}}\)^{\b_0} \Log{\tim^2 \over 2\pi \s}  \ ,
\label{pow1}
\ee where we omitted the constant term on the right hand side
since it can be easily removed by the finite renormalization.
The expression (\ref{pow1}) shows explicitly that the
instantons yield the powerlike corrections to the perturbative result, which
is expected from a general consideration.
Indeed, the powerlike behavior of the
nonperturbative corrections to the high energy QCD processes  may be
obtained from the renormalon analysis (see, {\it e.g.}, \cc{ren}).

It is instructive to express the result (\ref{pow1}) in terms of the mean
instanton size $\bar \r $ and the instanton density $\bar n$ calculated
directly from the distribution (\ref{dist1}), which read respectively
\be
\bar  \r = {\G(\b_0/2 - 3/2) \over \G(\b_0/2 - 2)} {1 \over \sqrt{2 \pi \s} } \ , \ee
\be
\bar n = {C_{N_{c}} \G (\b_0/2 - 2) \over 2} \(2\pi \over \a_S(\bar \r) \)^{2N_c}
\({\L_{QCD} \over \sqrt{2\pi \s}}\)^{\b_0} (2\pi \s)^2 \ .
\ee
Numerically, these quantities correspond to the usual mean size $\r_0$ and
density $n_0$ of the instanton liquid model, which parametrize the
deltalike approximation to the distribution (\ref{dist1}) (this approach,
which was proposed about two decades ago \cc{ilm}, now gets a strong support in
lattice data \cc{lat}):
\be
dn(\r) = n_0 \d(\r-\r_0) d\r \ , \label{de1}
\ee where the parameters were estimated in \cc{rev}: \be n_0 \approx 1 fm^{-4} \ \ ,
\ \ \r_0 \approx 1/3 fm \ . \label{param} \ee

To compare the instanton induced and perturbative parts, which are, respectively,
\be \left\{
\begin{array}
[c]{l}
w_I(\g) = 1 - K \ h(\g) \pi^2 \bar n \bar \r^4  \[\Log{\m^2 \over \l^2} - \Log {{\bar \r}^2\m^2} \] \\
W_P(\g) = 1 - {\a_S(\mu) \over 2 \pi} C_F h(\g) \Log {\m^2 \over \l^2} \ , \end{array}
\right. \label{kap2}
\ee where
\be
K = {\G(\b_0/2) [\G (\b_0/2-2)]^3 \over 2 \ [\G(\b_0/2-3/2)]^4} \approx
0.74 \ ,
\ee
we assume that the factorization scale $\m$ (which divides the soft and hard regions of momenta
in the factorized quark form factor) is of order of the inverse instanton size $\m \approx \bar \r^{-1}
\approx 0.6 \ GeV$.
Then we write the total leading order contribution to the Wilson loop expectation value in the
form \be
W(\g, \r_0\l) = 1 + {\a_S (\bar \r^{-1}) \over 2 \pi} C_F h(\g) \Log{ \( \bar \r^2\l^2 \)}
\(1 + K \ {S_0 \pi^2 \bar n \bar \r^4 \over  C_F}  \) \ , \label{kap4}\ee
where $S_0 = {8\pi^2 \over g^2(\bar \r^{-1})} \approx 10$
is the ``classical enhancement'' factor with the renormalized
coupling constant $g (\m)$ at the energy scale $\m \approx \bar \r^{-1}$.
The ratio of the instanton correction to the perturbative leading term
is given by
\be
{\hbox{\small{\sf Instanton}} \over \hbox{\small{\sf 1-Loop Perturb.}}} =
 K \ {S_0 \pi^2 \bar n \bar \r^4 \over  C_F} \approx 0.5 \ , \label{main2} \ee
which is estimated using the conventional value for the packing fraction
\cc{ilm}
\be \pi^2 \bar n \bar \r^4 \approx 0.1 \ . \ee
One can see using the main formula (\ref{pow1}) that
the strong power suppression of the instanton part
is partially compensated by the large factor $S_0^{2N_c+1}$.
This means that at the energy scale of order of $\bar \r^{-1}$ the magnitude of
the instanton induced effects is comparable to the leading perturbative
part, and must be taken into account as well.

Let us estimate now the two-loop perturbative contribution which may be
equally important at the chosen low
scale, where the strong coupling constant is not small enough [$\a_S (0.6 \ GeV) \approx
0.5$]. The two-loop result has been obtained in \cc{kr1} and in the limits
of small and large angles is given by
\be W^{(2)}(\g) \approx\left\{
\begin{array}
[c]{l}
\(\a_S \over \pi\)^2 C_F N_c {\g^2 \over3} \
\[{11 \over 48} \Log^2 {\m^2 \over \l^2} - \({3 \over 8} - {\pi^2 \over 12}
+ {67 \over 72} \) \Log {\m^2 \over \l^2}\] \qquad\mathrm{for\ small}\ \g ,\\
\({\a_S \over \pi} \)^2 C_F N_c \g \ \[{11 \over 48} \Log^2 {\m^2 \over \l^2} -
\( {67 \over 72} - {\pi^2 \over 24}\) \Log {\m^2 \over \l^2}\] \qquad\qquad\ \mathrm{for\ large}\ \g .
\end{array}
\right.
\label{twl1} \ee
From these expressions we get that the ratio of the instanton contribution
to the perturbative two-loop term is approximately equal to 1.4 and varies slightly (only
about $10\%$) with changing of $\g$ over whole allowed kinematical interval.
Thus, the complete consideration of the quark form factor at the low momentum scale must include
both the two-loop perturbative part and the leading order instanton one,
which appear to be of the same order of magnitude.

\vspace{1cm}

\section{Exponentiation of perturbative and instanton corrections}
\noindent
Expressions (\ref{kap4}, \ref{twl1}) define the first terms of the Wilson loop expansion in gauge
fields. On the basis of the exponentiation
theorem \cc{ExpTheo} for the non-Abelian path-ordered exponentials
it is well known that perturbative corrections to the Sudakov form factor are
exponentiated to high orders in the QCD coupling constant. The theorem
states that the contour average $W_{P}(C)$ can be expressed as
\be
W_{P}(C)=\exp{\[ \sum_{n=1}^{\infty}\(\frac{\a_S}{\pi}\)^n\sum_{W\in W(n)} C_n(W)F_n(W) \]},
\label{Exp}\ee
where summation in the exponential is over all diagrams $W$ of the set $W(n)$ of
the two-particle irreducible contour averages of $n$th order of the perturbative expansion.
The coefficients $C_n(W)\propto C_F N_c^{n-1}$ are the ``maximally non-Abelian'' parts
of the color factor corresponding to the contribution coming from a diagram $W$ to the total
expression (\ref{Exp}) in the contour gauge, and the factor $F_n(W)$ is the contour integral presented
in the expression for $W$.
This means that the essential diagrams are only those, which do not contain the lower-order
contributions as subgraphs and, as a result, the higher-order terms are non-Abelian.

Let us now demonstrate how the single instanton contribution is exponentiated in the small instanton
density parameter, treating the instanton vacuum as a dilute medium \cite{CDG78}.
The gauge field is taken to be the sum of individual instanton fields
in the singular gauge (\ref{if1}, \ref{SingG}), with their centers at the points $z_{j}$'s.
In this gauge, the instanton fields  fall off rapidly at
infinity, so the instantons may be considered individually in their effect
on the loop. Moreover, the contribution of infinitely distant
parts of the contour may be neglected and only those instantons will influence the loop integral,
which occupy regions of space-time intersecting with the
quark trajectories. Since the parametrization of the
loop integral along rays of the angle plays the role of the proper time,
a time-ordered series of instantons arises and has an effect on the Wilson loop.
Thus, the contribution of $n$ instantons to the loop integral $W_{I}(\g)$ can
be written in the dilute approximation as
\be
W^{(n)}_I(\g)=\Tr \( U^1 U^2...U^{n}U^{n\dagger}...U^{2\dagger }U^{1\dagger }\) ,
\ee
where the ordered line integrals $U_{i}$'s
\begin{eqnarray*}
U^{j}(\g) &=&T\left\{
\exp{\(ig\int_0^{\infty}\! d\s\ v_1^\m A_\m(v_1\s-z_j)\)}
\exp{\(ig\int^0_{-\infty} \! d\t \ v_2^\m A_\m(v_2\t-z_j)\)}
\right\}
\end{eqnarray*}
are associated with individual instantons with the positions $z_{j}$'s.
Because of the wide separation of the instantons in the dilute phase
and rapid falloff of fields in the singular gauge, the upper and lower limits of the
line integrals are extended to infinity.
The line integrals $U^{i\dagger}$'s take into account the part of the contour that goes at infinity
from $+\infty$ back to $-\infty$
and in the singular gauge $U^{i\dagger}=1$.
For $U^{j}(U^{j\dagger })$, the integral is taken over the increasing (decreasing) time piece of the loop.

Then, the expression is simplified when averaging over the gauge orientations
of instantons. The averaging is reduced to substitution of $U^{j}$ by
$g_{j}U^{j}g_{j}^{-1}$, where $g_{j}$ is an element of colour group, and independent integration
of each $g_{j}$ over the properly normalized group measure is performed. Under this
averaging one gets
\be
U^{n}U^{n\dagger }\rightarrow \frac{1}{N_{c}} \Tr \( U^{n}U^{n\dagger }\),
\ee
which is just the single instanton contribution $w_I^{(n)}(\g)$ as it is given by
Eqs. (\ref{wI1}, \ref{it1}).
But then if the averaging is done in the inverse order, from $n$ down to $1$,
the entire loop integral collapses to a product of traces
\be
W^{(n)}_I(\g)\rightarrow\lim_{n\to \infty }\prod_{j=1}^{n}
w_I^{(j)}(\g).
\ee
Since the individual instantons are considered to be decoupled in the dilute medium,
the total multiple instanton
contribution to the vacuum average of the Wilson operator simply exponentiates the all-order
single instanton term $w_I(\g)$ in (\ref{it1}), and one has
\begin{eqnarray}
W_I(\g)&=&\lim_{n \to \infty }\left\{ 1+\frac{1}{n} w_I(\g) \right\} ^{n}
=\exp [w_I(\g)].
\label{WDYinst}
\end{eqnarray}

Thus, we have proved that in the dilute regime, the full instanton contribution to
the quark form factor is given by the exponent of the all-order single instanton result.
The exponentiation arises due to taking into account the many-instanton configurations
effect. As it is well known, in QED there occurs the exponentiation of the one-loop result
due to Abelian character of the theory. In the instanton case, the analogous result takes place since
instantons belong to the $SU(2)$ subgroup of the $SU(3)$ color group and the path-ordered
exponents coincide with the ordinary ones. Exponentiation applied to the lowest order terms
(\ref{kap4}, \ref{twl1}) leads to vanishing the Wilson loop as a power of an IR cutoff $\lambda$.
Then, the additional instanton contributions suggest that the Wilson loop decreases
in the instanton medium faster with $\l$ than in purely perturbative theory.

\section{Conclusion}
\noindent
We have calculated the instanton contribution to the soft part of
the quark form factor, described in terms of the vacuum
expectation value of the Wilson loop for the contour of a special form (\ref{it1}, \ref{it2}).
Further analysis of high energy behavior
of this form factor must take into account the contributions of the hard and
collinear parts as well. In order to do this, we would need to ``free'' the
scale $\m$ and consider it as a factorization parameter lying between the
hard scale $Q$ and the IR cutoff $\l$ ($\l < \m < Q$). Then we would find
that the instantonic part, being UV finite due to the finite instanton size,
does not contain an explicit dependence on the UV cutoff $\m$, and therefore does not contribute to the
cusp anomalous dimension. However, this issue is out of scope of the present
paper, and we did not address this question here.

The situation may be
changed essentially in some other hadronic hard processes.
For example, the Wilson path for (in)elastic high energy
quark-quark scattering consists of two infinite straight lines lying in two
different planes, separated by the finite transverse distance \cc{nach, sh1}. The
calculation of the Wilson operator for this contour reduces
formally to the evaluation of the path integrals analogous to the ones
considered above. The only difference is that the sides of these contours
would belong to the different planes. The transverse distance between them provides
then an UV cutoff for the integrals, both for perturbative and instanton
fields, and the parameter of the dimensional regularization $\l$ becomes the
only (IR) cutoff for them. Therefore, the corresponding anomalous dimension would contain the
instanton induced part, and the instantons would play a nontrivial role in
the solution of the corresponding renormalization group equations. This topic
will be a subject of the forthcoming study.

Moreover, in the instanton calculations of the high energy parton scattering processes \cc{sh1},
the integrals over the instanton field can be
computed in all orders by virtue of the infinite length of lines that form
the integration path specific for this process.
The latter fact leads to the trivial factorization of the
$\g$-angle dependence from the corresponding Wilson integral, and extract
the instanton contribution into the numerically calculable form factors.
In our case, one deals with the cusp, rather than cross singularities.
The all-order calculation given formally by Eqs.(\ref{it1}, \ref{it2})
is quite complicated for the contour with cusp,
so we postpone the solution of this task.

To summarize, within the instanton vacuum model
we have developed an approach which allows one to calculate the
nonperturbative contributions to the Wilson integrals over the infinite
contour with a cusp that represent, {\it e. g.,} the classical trajectories of
partons participating in hard collisions.
We have proved that in the dilute regime, the full instanton contribution to
the quark form factor is given by the exponentiated all-order single instanton result, see (\ref{WDYinst}).
In the weak-field limit, the instanton contribution to the soft part of the
color singlet quark form factor is found explicitly
in terms of the instanton profile function in the singular gauge.
It is shown that the instanton induced effects are of a power type (\ref{pow1}), but
nevertheless they are comparable in magnitude
to the perturbative ones at the scale of order of the inverse average size of the
instanton in the instanton vacuum, see (\ref{kap4}).

\section{ Acknowledgments}
\noindent
We are grateful to D. Antonov, N.I. Kochelev, S.V.  Mikhailov, and S. Tafat.
This work is partially supported by RFBR (Grant Nos. 02-02-16194, 01-02-16431) and INTAS
(Grant No. 00-00-366). The work of I. Ch. was also supported by RFBR Grant
No. 00-15-96577. He is grateful to the Abdus Salam ICTP in Trieste for
the kind invitation and hospitality, where a part of this work was done.
\vspace{1.5cm}

\section{ Appendix}

\noindent
Below we write down some useful formulas, that have been used in the calculations.
The dimensionally regularized free propagator in the Euclidean space reads $(n=4-2\ve)$:
$$ \prop(z; \x) =  \l^{4-n} \int \! \dk \ex^{-ikz}
\({\d_{\m\n} \over k^2 } - \x {k_\m k_\n  \over k^4} \) =
$$
$$
= {(\l^2 \pi)^{(2-n/2)} \over 4\pi^2} \G(n/2-1) \(
\d_{\m\n}{(1-\x/2) \over z^{2(n/2-1)}} - \x z_\m z_\n {(n/2-1) \over
z^{2(n/2)}}\) \ . \eqno(1a)$$
The form factors $d_{1,2}(u)$ defined in Eq. (\ref{st1}) are given by ($u =z^2$ )
$$
d_{1}(u)={(\l^2 \pi)^{(2-n/2)} \over 16\pi^2} {\G(n/2-2) \over u^{(n/2-2)}}\ , \ \
 d_{2}(u)=\x d_{1}(u) \
. \eqno(1b)$$
 The partial derivatives in terms of the derivative with respect to the interval
$u$ in $n$-dimensional space-time
$$ \pd_z^2 = 2 n \pd_u + 4 u \pd^2_u \ , \eqno(2a)$$
$$ \pd_\m \pd_\n = 2 \d_{\m\n} \pd_u + 4 z_\m z_\n \pd^2_u  \ . \eqno(2b)$$
Scalar products in the Euclidean space $$ v^1_\m v^2_\n\d_{\m\n} = \cos \g \ \ ,
\ \ v^1_\m v^2_\n z_\m z_\n = z^2 \ \cos
\g + s t \ \sin^2 \g \ . \eqno(3)$$
For calculations of the perturbative part we used the following integrals:
$$
\int_0^{\infty} \! ds dt \ \ex^{-\a (s^2 + t^2 + 2 st \ \hbox{\footnotesize
cos} \g)} = {1 \over 2 \a}{\g \over \sin \g} \  \eqno(4)$$
and
$$ \int_0^{\infty} \! ds dt \ st \ \ex^{-\a (s^2 + t^2 + 2 st \
\hbox{\footnotesize cos} \g)} =  {1 - \g \cot \g \over 4 \a^2 \sin^2 \g} \ .
\eqno(5)$$
Therefore, one has for a function $d(u)$:
$$ d_i^{\{k\}}(u) = (-)^k \int_0^{\infty} \! d\a \a^k \ex^{-\a
u} \bar d_i(\a) \  \eqno(6)$$ and gets
$$\int_0^{\infty} \! ds dt \ d' (u) = - {\g \over 2 \sin \g} d(0) \ , \eqno(7a)$$
$$\int_0^{\infty} \! ds dt \ u d'' (u) = {\g \over 2 \sin \g} d(0) \ , \eqno(7b)$$
$$ \int_0^{\infty} \! ds dt \ st d'' (u) = {1 - \g \cot \g \over 4 \sin^2
\g} d(0) \ . \eqno(7c)$$

\begin{figure}
\begin{flushleft}
\includegraphics[height=20cm, width=1.55\textwidth]{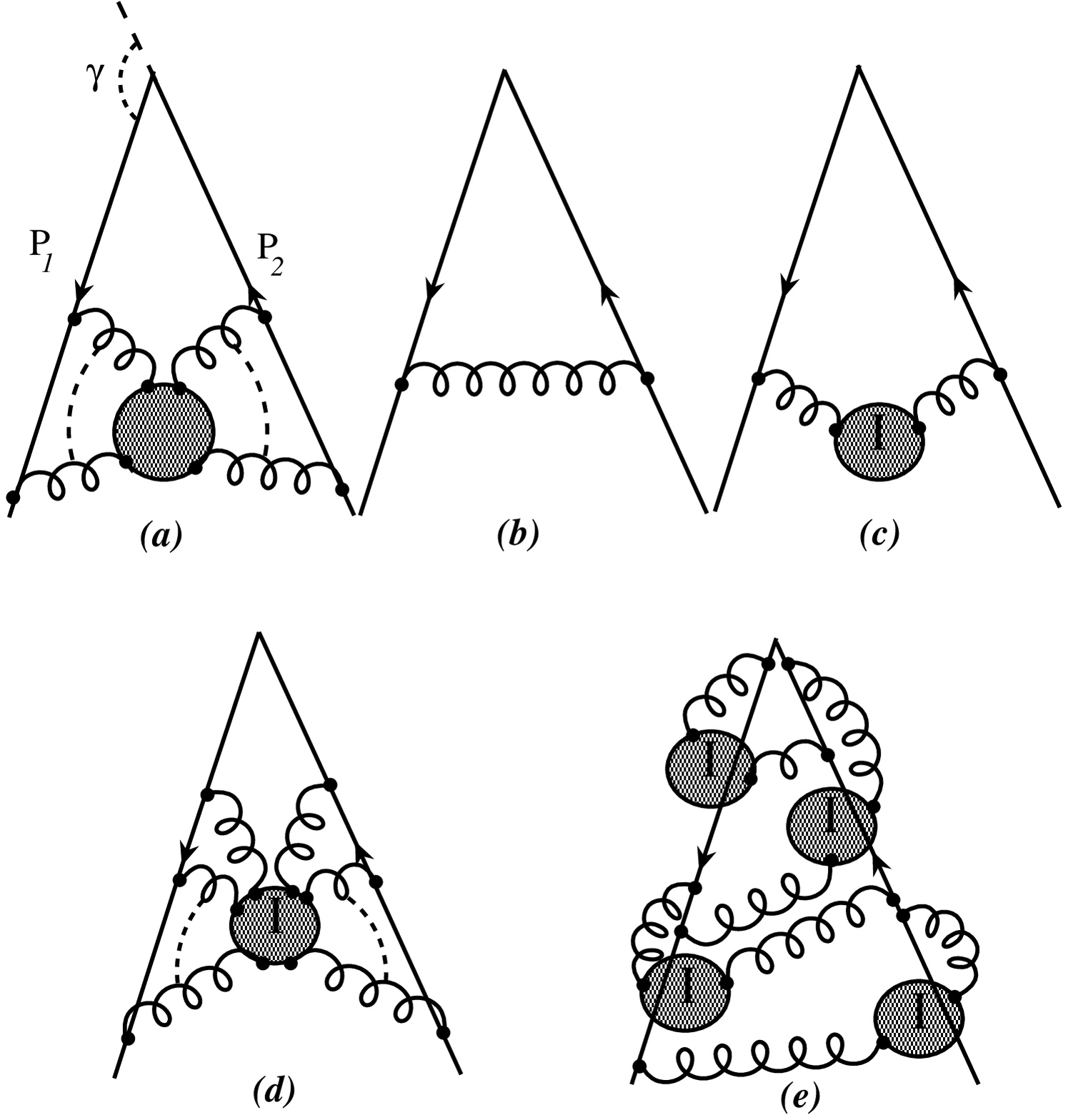}
\end{flushleft}
\noindent
{\bf Fig. 1:} The notations for the quark momenta and the total
cusp-dependent part of the Wilson loop integral for the quark form factor (a);
the leading order contributions of the  perturbative (b) and
nonperturbative (single-instanton) (c) fields;
 (d) the all-order single instanton result; (e) the exponentiation of the
single instanton result.
\end{figure}


\vspace{1cm}

\begin{thebibliography}{99}
\bibitem{pol} Yu. M. Makeenko, A. A. Migdal, {\it Phys. Lett. B} 88 (1979) 135; {\it Nucl. Phys.
B} 188 (1981) 269;  A. Polyakov, {\it Gauge Fields and Strings}, Harwood,
1987.

\bibitem{hard}A. Bassetto, M. Ciafaloni, G. Marchesini, {\it
Phys. Reports} 100 (1983) 201; G. Korchemsky, {\it Phys. Lett. B} 220 (1989)
629; G. Korchemsky, G. Sterman, {\it Nucl. Phys. B}
437 (1995) 415;  S. Tafat, {\it JHEP} 05 (2001) 004;
G. Korchemsky, A. Radyushkin, {\it Phys. Lett. B} 171 (1986) 459.

\bibitem{nach} O. Nachtmann, {\it Ann. Phys.} (N.Y.) 209 (1991) 436.

\bibitem{korh} G. Korchemsky,  A. Radyushkin, {\it Phys. Lett. B} 279 (1992) 359.

\bibitem{rev} T. Sch\"afer, E. V. Shuryak, {\it Rev. Mod. Phys.} 70 (1998)
323.

\bibitem{lat} D. A. Smith, M. J. Teper, {\it Phys. Rev. D} 58 (1998) 014505;
J. W. Negele, {\it Nucl. Phys. B (Proc. Suppl.)} 73 (1999) 92.

\bibitem{ring} S. Moch, A. Ringwald, F. Schrempp, {\it Nucl. Phys. B} 507 (1997) 134.

\bibitem{sh1} E. Shuryak, I. Zahed, {\it Phys. Rev. D} 62 (2000) 085014;
M. Nowak, E. Shuryak, I. Zahed,  {\it Phys. Rev. D} 64 (2001) 034008.

\bibitem{DEMM99} A.E. Dorokhov, S.V. Esaibegyan, S.V. Mikhailov,
{\it Phys. Rev. D} {56} (1997) 4062;
A.E. Dorokhov, S.V. Esaibegyan, A. E. Maximov, S.V. Mikhailov, {\it Eur. Phys. J. C} {13 }
(2000) 331, hep-ph/9903450.

\bibitem{kr217} G. Korchemsky, {\it Phys. Lett. B} 217 (1989) 330.

\bibitem{ao} V. Dotsenko, S. Vergeles, {\it Nucl. Phys. B} 169 (1980) 527;
A. Polyakov, {\it Nucl. Phys. B} 164 (1980) 171.


\bibitem{bra} R. A. Brandt, F. Neri, M.-A. Sato, {\it Phys. Rev. D} 24 (1981)
879;  R. A. Brandt, A. Gocksch, M. A. Sato, F. Neri, {\it Phys. Rev. D} 26 (1982) 3611.

\bibitem{kr325} G. Korchemsky, {\it Phys. Lett. B} 325 (1994) 459.

\bibitem{kr1} G. Korchemsky, A. Radyushkin, {\it Nucl. Phys. B} 283 (1987) 342.

\bibitem{meg} E. Meggiolaro, {\it Phys. Rev. D} 53 (1996) 3835.

\bibitem{hard0} N. S. Craigie, H. Dorn, {\it Nucl. Phys. B} 185 (1981);
D. Knauss, K. Scharnhorst, {\it Ann. der Phys.} {( Leipzig)} 41 (1984)
331.

\bibitem{ExpTheo} J.G.M. Gatheral, {\it Phys. Lett. B} 133 (1983) 90;
J. Frenkel, J.C. Taylor, {\it Nucl. Phys. B} 246 (1984) 231.

\bibitem{shdist} E. Shuryak, hep-ph/9909458.

\bibitem{ren} M. Beneke, {\it Phys. Reports} 317 (1999) 1.

\bibitem{ilm} E. Shuryak, {\it Nucl. Phys. B} 203 (1982) 93, 116, 140.

\bibitem{CDG78}  C. G. Callan, R. Dashen, D. J. Gross, {\it Phys. Rev. D} 17 (1978) 2717.






\end{thebibliography}
\end{document}